\documentclass[twoside,twocolumn]{article}
\usepackage[super,sort&compress,comma]{natbib} 
\usepackage{mhchem}
\usepackage{times,mathptmx}
\usepackage{sectsty}
\usepackage{balance} 

\usepackage{graphicx} 
\usepackage{lastpage}
\usepackage[format=plain,justification=raggedright,singlelinecheck=false,font=small,labelfont=bf,labelsep=space]{caption} 
\usepackage{fancyhdr}
\usepackage[T1]{fontenc} 
\usepackage[latin2]{inputenc}
\usepackage{amssymb}
\usepackage{amsthm}
\usepackage{color}

\usepackage[
   separate-uncertainty  = true,
   mode = math,
   multi-part-units      = brackets,
]{siunitx} 
\usepackage{hyperref}
\hypersetup{
    colorlinks=true, 
    linkcolor=blue,  
    citecolor=blue,  
}
\usepackage[all]{hypcap}  

\renewcommand{\_}[1]{{}_{\mathrm{#1}}}

\begin{document}

\renewcommand{\thefootnote}{\fnsymbol{footnote}}
\renewcommand\footnoterule{\vspace*{1pt}%
\hrule width 3.4in height 0.4pt \vspace*{5pt}} 

\makeatletter 
\def\subsubsection{\@startsection{subsubsection}{3}{10pt}{-1.25ex plus -1ex minus -.1ex}{0ex plus 0ex}{\normalsize\bf}} 
\def\paragraph{\@startsection{paragraph}{4}{10pt}{-1.25ex plus -1ex minus -.1ex}{0ex plus 0ex}{\normalsize\textit}} 
\renewcommand\@biblabel[1]{#1}            
\renewcommand\@makefntext[1]%
{\noindent\makebox[0pt][r]{\@thefnmark\,}#1}
\makeatother 
\renewcommand{\figurename}{\small{Fig.}~}
\sectionfont{\large}
\subsectionfont{\normalsize} 

\renewcommand{\headrulewidth}{1pt} 
\renewcommand{\footrulewidth}{1pt}
\setlength{\arrayrulewidth}{1pt}
\setlength{\columnsep}{6.5mm}
\setlength\bibsep{1pt}

\fancyhead{}
\renewcommand{\headrulewidth}{1pt} 
\renewcommand{\footrulewidth}{1pt}
\setlength{\arrayrulewidth}{1pt}
\setlength{\columnsep}{6.5mm}
\setlength\bibsep{1pt}

\twocolumn[
  \begin{@twocolumnfalse}
\noindent\LARGE{\textbf{Surface-confined 2D polymerization of a brominated copper-tetra\-phenylporphyrin on Au(111)}} 
\vspace{0.6cm}

\noindent\large{\textbf{Lars~Smykalla,$^{\ast}$\textit{$^{a}$} Pavel~Shukrynau,\textit{$^{a}$} Marcus~Korb,\textit{$^{b}$} Heinrich~Lang,\textit{$^{b}$} and Michael~Hietschold\textit{$^{a}$}}}\vspace{0.5cm}

\vspace{0.6cm}

\noindent \normalsize{
A coupling-limited approach for the Ullmann reaction-like on-surface synthesis of a two-dimensional covalent organic network starting from a halogenated metallo-porphyrin is demonstrated. Copper-octabromo-tetraphenylporphyrin molecules can diffuse and self-assemble when adsorbed on the inert Au(111) surface. Splitting-off of bromine atoms bonded at the macrocyclic core of the porphyrin starts at room temperature after the deposition and is monitored by X-ray photoelectron spectroscopy for different annealing steps. Direct coupling between the reactive carbon sites of the molecules is, however, hindered by the molecular shape. This leads initially to an ordered non-covalently interconnected supramolecular structure. Further heating to \SI{300}{\celsius} and an additional hydrogen dissociation step is required to link the molecular macrocycles via a phenyl group and form large ordered polymeric networks. This approach leads to a close-packed covalently bonded network of overall good quality. The structures are characterized using scanning tunneling microscopy. Different kinds of lattice defects and, furthermore, the impact of polymerization on the HOMO--LUMO gap are discussed. Density functional theory calculations corroborate the interpretations and give further insight into the adsorption of the debrominated molecule on the surface and the geometry and coupling reaction of the polymeric structure.}
\vspace{0.5cm}
 \end{@twocolumnfalse}
  ]

\section{Introduction}

\footnotetext{\textit{$^{a}$~Technische Universität Chemnitz, Institute of Physics, Solid Surfaces Analysis Group, D-09107 Chemnitz, Germany; E-mail: lars.smykalla@physik.tu-chemnitz.de; hietschold@physik.tu-chemnitz.de}}
\footnotetext{\textit{$^{b}$~Technische Universität Chemnitz, Institute of Chemistry, Inorganic Chemistry, D-09107 Chemnitz, Germany}}

The formation of molecular nano-structures on surfaces through self-assembly of building blocks which are held together only by weak intermolecular interactions has the inherent problems that these arrangements are fragile and intermolecular charge transport is limited. A solution to this is a controlled interconnection of the molecules on the surface by robust and irreversible covalent bonds. 
Covalent linking of small organic molecules directly on the surface allows also for the engineering of manifold extended 2D materials which are otherwise not accessible by classical solution-based polymerization. Especially, the prospect to fabricate $\pi$-conjugated graphene-like nano-sheets of controlled dimensions is of great interest.\cite{Bronner2014,Fan2013,Lipton-Duffin2009} Therefore, the on-surface synthesis of 1D and 2D polymers has attracted considerable attention over the recent years.\cite{Franc2011,Garah2013,Bjoerk2014,Gourdon2008,Sakamoto2009}

One major approach for surface-confined polymerization is the halogen-based covalent self-assembly based on the Ullmann coupling reaction.\cite{Ullmann1901} Halogen atom substituents are split-off a monomer and, thus, unsaturated carbon atoms at predefined positions are created. Thereby, the influence of a catalyst (originally Cu) or the adsorption on a surface significantly lowers the energy barrier for dissociation. Subsequently, the resulting molecular radicals diffuse to fitting positions close to another to enable C--C coupling into covalent dimers, chains or networks. The topography and also electronic properties of the covalent organic framework are determined by the monomer used and its halogen substitution pattern. Also, the properties of the substrate surface, like the reactivity and the energy barriers for diffusion and rotation, play an important role in this kind of synthesis.\cite{Bieri2010, Gutzler2009, Walch2010}

The polymerization is often observed to not occur directly after the dehalogenation, rather an additional activation step is required, which is achieved typically photochemically\cite{Palma2014} or by heating. Also, it was demonstrated that the reaction steps, which are halogen dissociation, movement of the molecules and C--C coupling, could even be individually and locally performed by voltage pulses and manipulation with the tip of a scanning tunneling microscope.\cite{Hla2000}

It was reported that dehalogenation started already at reduced temperature and is nearly finished at room temperature\cite{Schmid2011,Cardenas2013,Eder2013}, especially on the reactive Cu surfaces. Afterwards, the radicals can be stabilized in organometallic structures by metal atoms, \textit{e.g.} Cu surface adatoms.\cite{Gutzler2009,Gutzler2014,Walch2010,Fan2013} For actual polymerization, additional heating often to high temperatures is then necessary to cleave the carbon--metal bonds and enable diffusion that subsequently leads to irreversible C--C coupling of the dehalogenated radicals.\cite{Cardenas2013,Eichhorn2014a} This diffusion-limited polymerization route, however, leads to the typical main problems of 2D polymers, which are small domain sizes\cite{Cardenas2013}, a high density of defects or undesired crumpled or dendritic networks without long-range order.\cite{Bieri2010,Gutzler2014,Eichhorn2014a}
Split-off halogen atoms which remain on the surface stabilized by chemisorption and the presence of metal adatoms can also hinder the formation of long-range ordered covalent networks\cite{Lipton-Duffin2009,Bieri2010}. Considering that a high mobility of molecules is a prerequisite for the formation of well-ordered structures, Cu(111) seems also not to be the ideal surface for halogen-based covalent self-assembly due to its relatively high diffusion barriers.\cite{Bjoerk2013}
The optimal conditions for large ordered 2D networks are a surface with lower diffusion barriers like Au(111) as well as a relatively large recombination barrier of the monomer radicals, which results in a coupling-limited polymerization process.\cite{Bjoerk2014,Lafferentz2012} This means, monomers are able to diffuse and rearrange to a greater extent before they finally irreversibly combine into polymeric networks. Notably, probably the first example of halogen-based, two-dimensionally covalent self-assembly on a surface was demonstrated on Au(111) by Grill and coworkers.\cite{Grill2007} However, higher temperatures to effectively initiate the reaction steps are necessary on Au compared to Cu surfaces, which could already lead to further decomposition of large monomers.

In general, scanning tunneling microscopy (STM) has been the main tool for exploring surface-confined polymerization. Although, despite the formation and structure of various 2D polymers were reported, only little attention was given to their electronic structure.\cite{Bronner2014}
Compared to covalently bonded networks of metal-free molecules, even more interesting transport properties are expected, when incorporated metal ions can be coupled. For applications in molecular spintronic devices, the spin bearing metal ions should be decoupled from the surface, which mostly excludes the metal-organic coordination assemblies linked by metal adatoms.\cite{Li2012,Haq2011,Iancu2006,Shi2009,Stepanow2008} A good candidate could be networks of metallo-tetraphenylporphyrin molecules\cite{Tanoue2014}, which additionally should have a larger mechanical robustness and stability on the surface compared to non-covalently bonded structures of ``magnetic'' molecules.
Nevertheless, the largest challenge currently still lies in improving the domain size and quality of the polymeric organic networks.
Previously, Grill and coworkers investigated a metal-free tetraphenylporphyrin with bromine at the para position of the phenyl groups which couples the molecules at these C atoms of the phenyl rings after the cleavage of the C--Br bonds and, consequently, forms networks with a low packing density.\cite{Grill2007}
In this paper, we report the steps for polymerization of copper-2,3,7,8,12,13,17,18-octabromo-5,10,15,20-tetraphenylporphyrin (CuTPPBr$_8$) monitored by a combination of STM and X-ray photoelectron spectroscopy (XPS) and supported by density functional theory calculations. The different bromine substitution results in a hindered direct C--C coupling and, thus, a new coupling scheme and close-packed covalently bonded network, which will be discussed in the following.
\newpage

\section{Experimental details}

\begin{figure}[tb]
\centering
\includegraphics[width=0.40\textwidth]{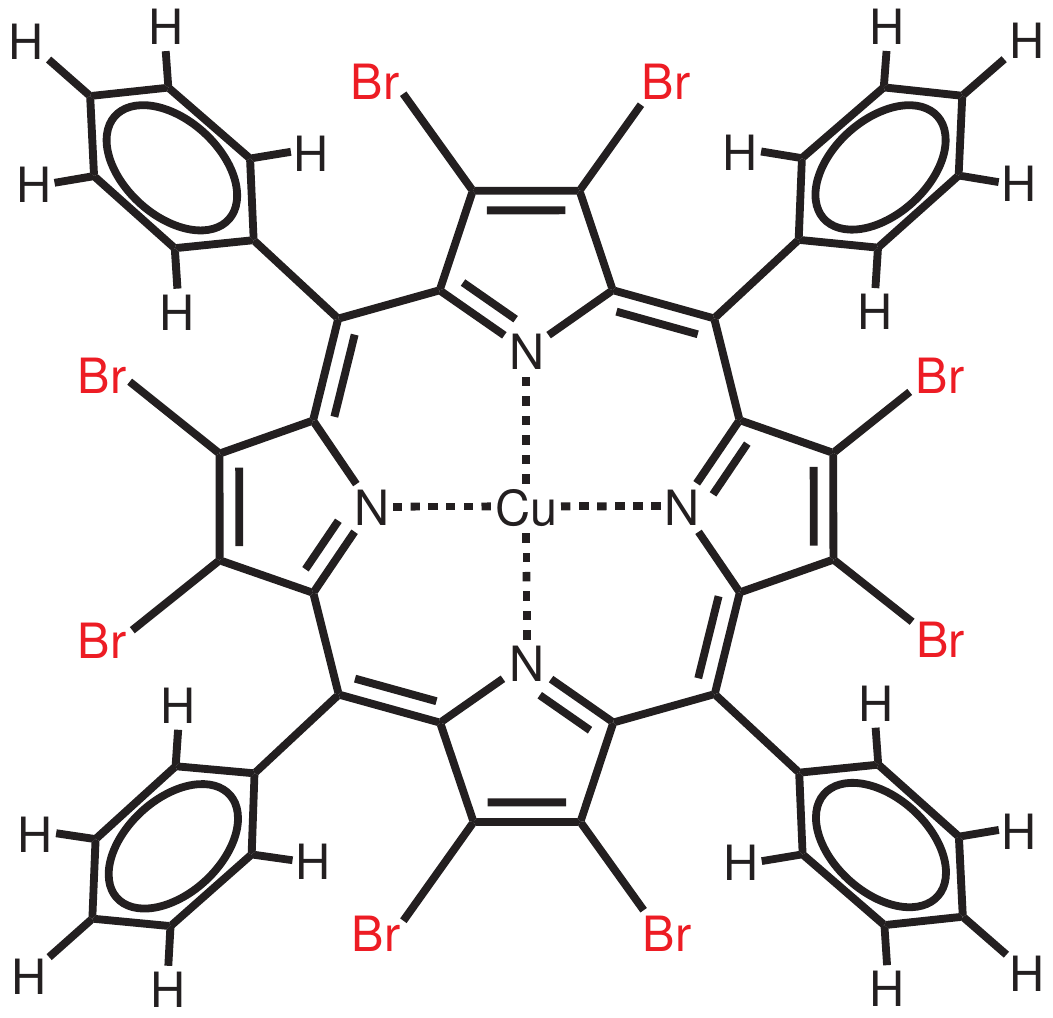}
\caption{Chemical structure of copper-2,3,7,8,12,13,17,18-octabromo-5,10,15,20-tetraphenylporphyrin (CuTPPBr$_8$)}
\label{fig:mol}
\end{figure}

A clean surface of the single crystal of Au(111) was prepared by multiple cycles of Ar$^+$-sputtering at an energy of \SI{500}{eV} and annealing to \SI{400}{\celsius} for \SI{1}{\hour}. 
CuTPPBr$_8$ was synthesized by a similar procedure as reported in Ref.~\citenum{Honda2010}. The chemical structure is shown in Fig.~\ref{fig:mol}. The molecules were deposited on the substrate by organic molecular beam epitaxy in ultra high vacuum (UHV). Before, the evaporant was first purified by heating to a temperature slightly below the sublimation temperature in UHV until the pressure dropped again. The molecules were then deposited at around \SI{350}{\celsius} on Au(111). The substrate was kept near room temperature during deposition.
Scanning tunneling microscopy experiments were performed with a variable temperature STM from Omicron in UHV. The base pressure in the UHV chamber was in the range of \SI{E-10}{mbar} during measurements. Electrochemically etched tungsten tips for the STM were cleaned by annealing in UHV. All measurements were done at room temperature and STM images were recorded in constant current mode with a tunneling current of \SI{100}{pA} if not specified otherwise. Positive bias corresponds to electron tunneling from the tip into empty states of the sample. STM images were processed with the WSxM software.\cite{Horcas2007}

X-ray photoelectron spectroscopy experiments were performed with synchrotron radiation at the Material Science beam line in Elettra (Trieste, Italy). A Phoibos photoelectron spectrometer was used. The excitation photon energy was for Br 3d \SI{180}{eV}, for C 1s \SI{400}{eV}, and for N 1s \SI{520}{eV}. The binding energies were corrected relative to the Au 4f$\_{7/2}$ peak of the substrate located at \SI{84}{eV}. The shapes of the core-level signals were fitted using Voigt peaks and a Shirley background. The samples were heated up to the temperature indicated in the figures, which was measured directly at the backside of the crystal, held at this temperature for \SI{10}{\minute} and then cooled down to room temperature prior to the XPS measurement.

Density Functional Theory (DFT) calculations were carried out using the grid-based projector augmented wave method.\cite{Larsen2009} The RPBE exchange-correlation functional with a pair-wise correction for the dispersion interaction [vdW(TS)]\cite{Tkatchenko2009} and the vdW-parameters for Au by Ruiz \textit{et al.} were used to include screening effects of the metal surface.\cite{Ruiz2012}. A grid spacing of $h = \SI{0.2}{\angstrom}$, and a dzp basis set of atomic orbitals for the wave functions were used. For the self-interaction of the $d$ states of the Cu atoms a Hubbard $U$ correction with a value of $U = \SI{5}{eV}$ was applied. Simulation cells with periodic boundary conditions in $x$ and $y$ and zero boundary in $z$ direction with a total vacuum of both sides of \SI{1.8}{nm} were used.
For the calculation of debrominated CuTPP on Au(111), a slab of 3 layers of Au was used, whereby the two lowest layers were fixed during optimization. The reconstruction of the Au(111) surface was not considered within the calculations. The structures were relaxed until all atomic forces converged. Only the $\Gamma$-point was used for the Brillouin zone sampling due to the relative large size of the cell.

\section{Results and discussion}
\subsection{XPS investigations}

\begin{figure}[!tb]
\centering
\includegraphics[width=0.48\textwidth]{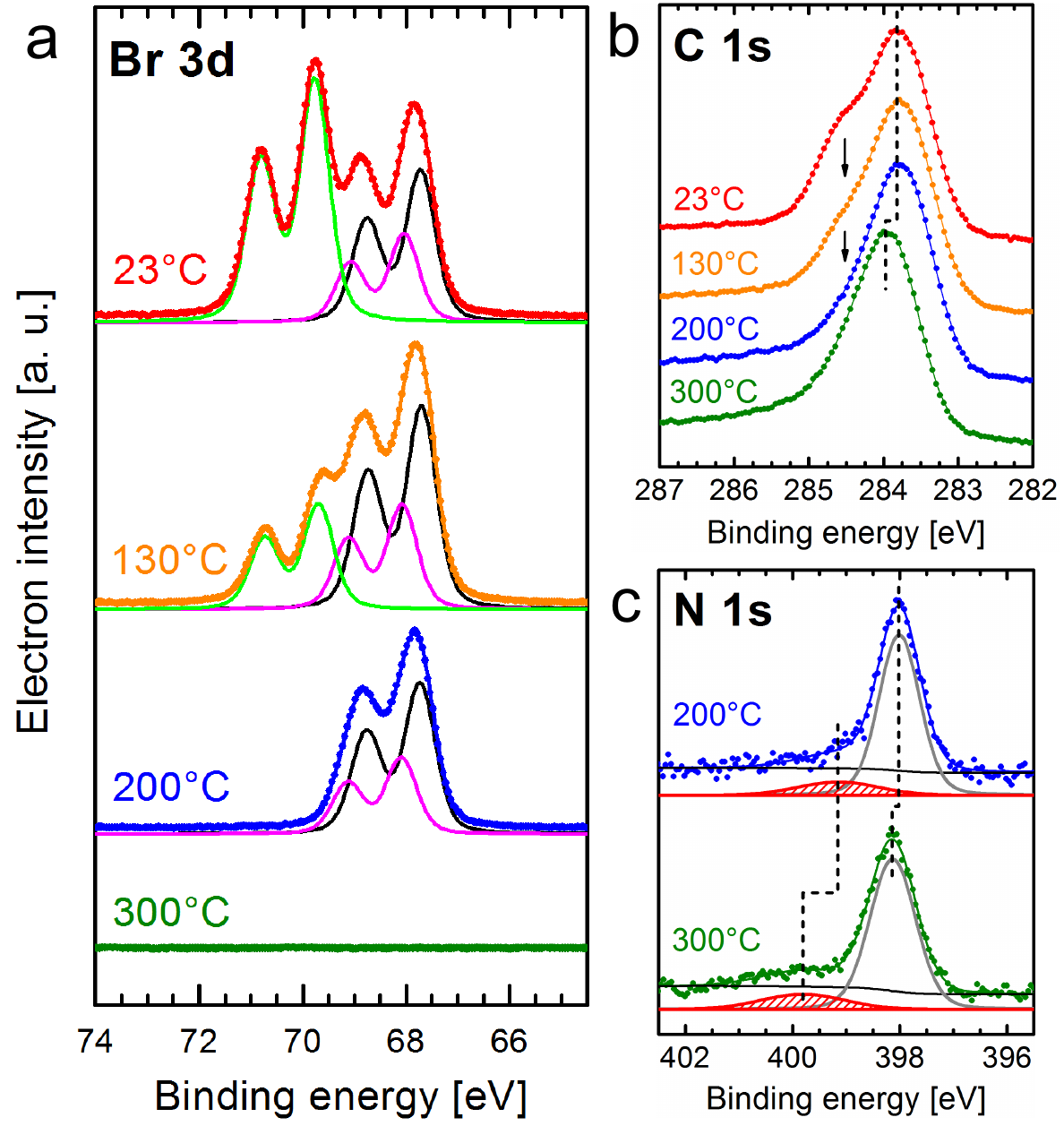}
\caption{Evolution of the XPS spectra of Br 3d (a), C 1s (b), and N 1s (c) for a molecular monolayer of CuTPPBr$_8$ on Au(111) compared between annealing steps to the indicated maximal temperature. In the Br 3d spectra, the background is subtracted.}
\label{fig:PES}
\end{figure}

The evolution of the XPS core-level spectra with increasing temperature is shown in Fig~\ref{fig:PES}. 
For Br 3d, the shape has to be fitted with three components if an identical width and spin-orbit splitting (\SI{1.04}{eV}) is used for the doublet peaks. The doublets with the larger Br 3d$_{5/2}$ peaks located at \SI{67.7}{eV} and \SI{68.1}{eV} (black and pink curves) are attributed to Br$_2$ \cite{Bulusheva2012} and Br chemisorbed on the surface.\cite{Cardenas2013,Gutzler2014,Wagner1991,Eichhorn2014a} The doublet at the higher binding energy with the Br 3d$_{5/2}$ peak at \SI{69.7}{eV} (green curve) was reported to correspond to bromine bonded to carbon\cite{Bulusheva2012, Cardenas2013, Eichhorn2014a}, \textit{i.e.} Br atoms in the intact CuTPPBr$_8$ molecule.
After deposition at room temperature, a high amount of dissociated bromine (doublets at lower binding energy) was found. It is to be expected that the Br atoms in CuTPPBr$_8$ which are in direct contact to the surface have a lower energy barrier for dissociation than the Br atoms which are bended away from the surface due to the saddle deformation of the molecule. It is also likely that some molecules already partially debrominated during the sublimation in the crucible or when arriving with high thermal energy on the sample surface.\cite{Grill2007,Eichhorn2014a}. The dissociated bromine atoms, which are chemisorbed on the Au surface, could then also recombine with the radical molecules or form Br$_2$.

After heating the sample to \SI{130}{\celsius} for \SI{10}{\minute}, the C--Br peaks were reduced while the other doublets slightly increased in intensity, which means that further molecules were debrominated. If the sample is annealed to a maximum of \SI{200}{\celsius}, the debromination of all molecules is finished and Br atoms or Br$_2$ remain adsorbed on the surface but also already start to desorb from Au(111). Finally, after heating to \SI{300}{\celsius}, no Br 3d signal was found anymore, \textit{i.e.} bromine completely disappeared from the Au(111) surface.
Furthermore, partial desorption (coverage reduced by 25\%) after annealing to \SI{300}{\celsius} is indicated by the increase of intensity of the XPS peaks of the Au substrate (Au 4f) which are attenuated by the molecular layer.
The temperature necessary for debromination on the relatively inert Au(111) is higher than those reported for other brominated molecules on Cu or Ag surfaces, where cleavage of the C--Br bonds was found to be complete at or even below room temperature, or at slightly elevated temperatures, respectively.\cite{Gutzler2014, Cardenas2013}
In studies on Au(111), debromination and C--C coupling for a tetraphenylporphyrin with a different bromination scheme than CuTPPBr$_8$ was, however, found in STM studies to occur only after heating to \SI{300}{\celsius}.\cite{Grill2007, Lafferentz2012}

The heating induced debromination can also be observed in the C 1s core level spectra [Fig.~\hyperref[fig:PES]{\ref*{fig:PES}(b)}]. A peak component at $\approx$ \SI{284.5}{eV} (arrows) shrinks with increasing temperature. We attribute this feature to carbon bonded to bromine.\cite{Lee2007a} However, this species overlaps with the peak for C--N at almost the same energy, which makes a non-ambiguous deconvolution of component intensities difficult. The shift of the C 1s (and also N 1s) signal to higher binding energy by \SI{0.2}{eV} after heating to \SI{300}{\celsius} is related to a shift of the Fermi level\cite{Bulusheva2012} due to the desorption of bromine from the surface and the polymerization of the molecules, which would cancel a possibly strong interaction of unsaturated carbon atoms with the surface. Annealing to \SI{400}{\celsius} leads to no further changes in the XPS signals.

Small peaks at higher binding energies compared to the main core-level peaks are often shake-up satellites due to an additional $\pi \rightarrow \pi^\star$ excitation. For N 1s  [Fig.~\hyperref[fig:PES]{\ref*{fig:PES}(c)}] with only a single component (N--Cu), the satellite can be clearly identified. Because the N orbitals contribute to the highest occupied (HOMO) as well as lowest unoccupied (LUMO) molecular orbital, the excitation energy can be correlated with the energy gap between the HOMO and LUMO.\cite{Haidu2014} However, the distance of the shake-up satellite to the corresponding core-level peak could be smaller than the actual HOMO--LUMO gap due to reorganization of the charge within the molecule upon photoexcitation and, thus, changed final state screening of the core hole.\cite{Rocco2008} The separation from the main N 1s peak at \SI{398}{eV} is \SI{1.2}{eV} for the completely debrominated molecules. Heating to \SI{300}{\celsius} leads, besides the $+\SI{0.2}{eV}$ shift of all peaks related to the molecule, to a larger shift of the satellite peak indicating a higher HOMO $\rightarrow$ LUMO excitation energy of \SI{1.6}{eV}. This means that by polymerization the band gap widened by $\approx$\SI{0.4}{eV}. Gas phase calculations using DFT [RPBE+$U$(\SI{5}{eV})+vdW(TS)] give a HOMO--LUMO gap of \SI{1.45}{eV} for CuTPPBr$_8$, \SI{0.86}{eV} for the debrominated molecule CuTPP, \SI{1.04}{eV} for an unsaturated CuTPP-polymer and \SI{1.58}{eV} for the H-saturated CuTPP-polymer. It should be noted that DFT generally underestimates energy gaps. On the other hand, the HOMO--LUMO gap is significantly reduced in the proximity of a surface compared to a thick layer\cite{Haidu2014} or the gas phase as calculated here, which leads to the seemingly good agreement for CuTPP and the H-saturated CuTPP-polymer with the experimental values.
After debromination the HOMO--LUMO gap is strongly reduced due to C atoms with unsaturated bonds. After C--C coupling of the CuTPP monomers the gap is increased by circa \SI{0.2}{eV} if the C atoms in the polymer would still be unsaturated. A value slightly higher than the energy gap of a single CuTPPBr$_8$ is found if the pyrrolic C atoms of the final polymer are fully saturated with the H atoms previously bonded at the phenyl groups. In recent studies, a trend of band-gap contraction with increasing sheet size of various 2D polymers was reported in theoretical gas phase calculations\cite{Gutzler2013, Cardenas2013} and scanning tunneling spectroscopy\cite{Chen2013} due to an extended $\pi$-conjugation. We propose that the observed opening of the HOMO--LUMO gap after C--C coupling is also due to the reduced hybridization of orbitals of the unsaturated carbon atoms with the Au surface. Furthermore, unlike completely $\pi$-conjugated networks such as graphene, the electron delocalization in the described CuTPP polymer is likely limited because of a large twist angle between the p-orbitals in the phenyl groups and in the porphyrin macrocycle, which separates the respective $\pi$-systems.\cite{Cardenas2013,Bronner2014}

\subsection{STM investigations of the molecular adlayer}

\begin{figure}[tb]
\centering
\includegraphics[width=0.48\textwidth]{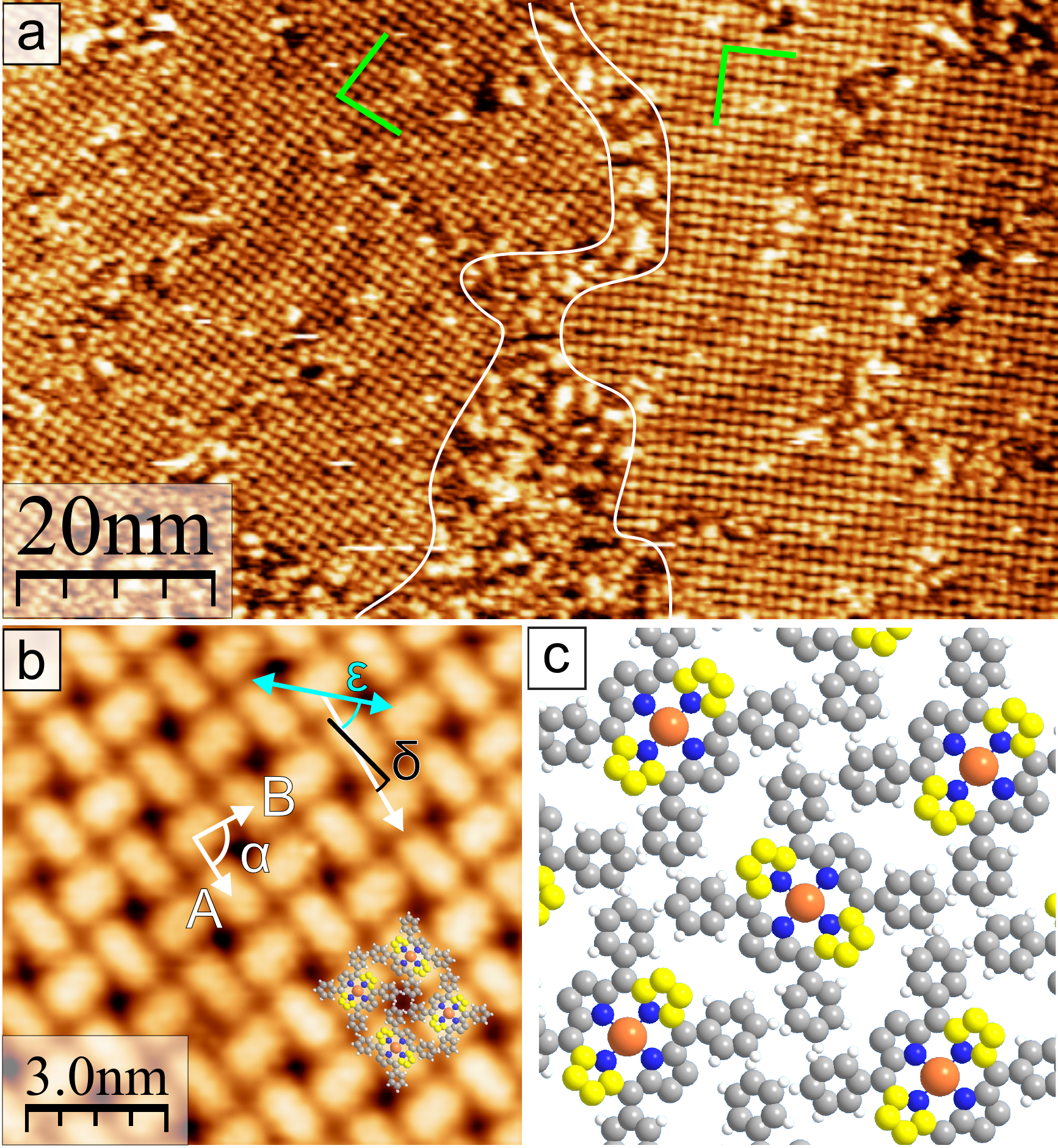}
\caption{(a) STM image after deposition of $\approx$ 1 ML of CuTPPBr$_8$ and annealing to max. \SI{200}{\celsius}. Two periodically ordered rotational domains and disordered areas in between are shown. The herring-bone reconstruction of the underlying Au(111) is visible through the large ordered areas. ($U = \SI{-0.7}{V}$) (b) Magnification of the adlayer structure of CuTPPBr$_8$ on Au(111) with elongated appearance of molecules at $U = \SI{-1}{V}$. The cyan arrow indicates the direction of an Au lattice vector. (c) Model of the molecular arrangement; Protruding C atoms of the saddle deformation are colored yellow.}
\label{fig:struc}
\end{figure}

\begin{figure*}[tb]
\centering
\includegraphics[width=1.0\textwidth]{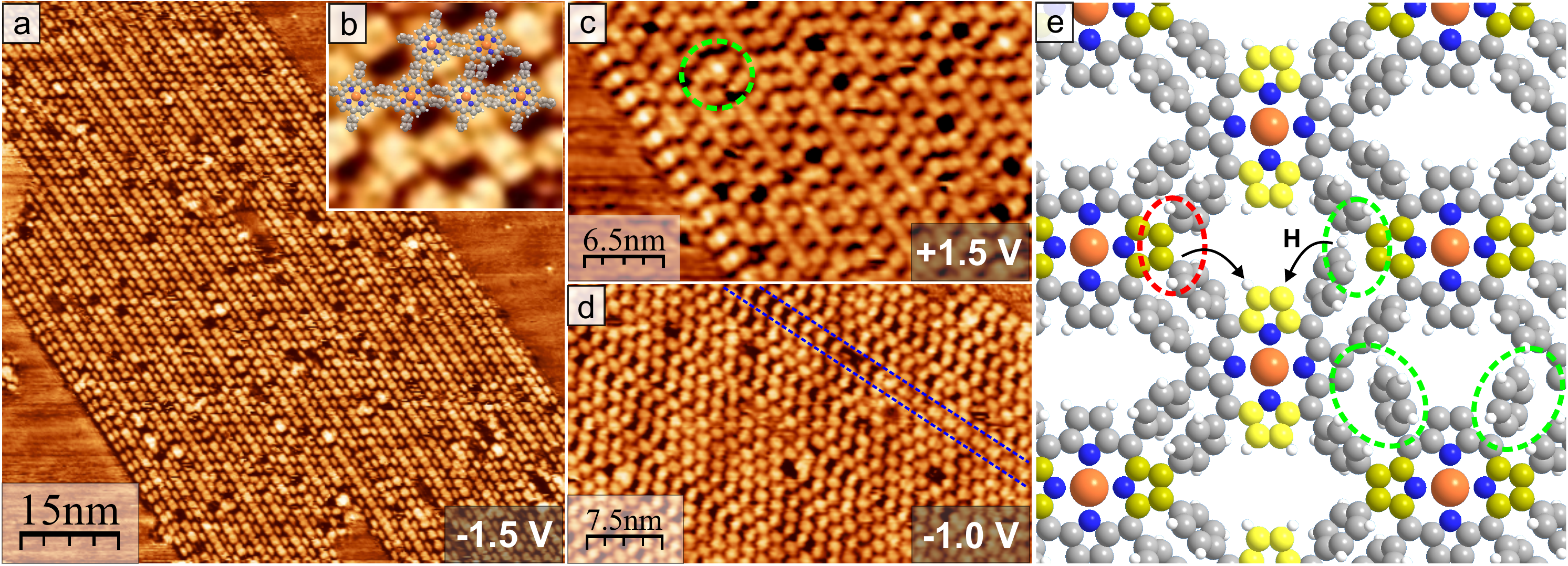}
\caption{(a) New structure of the alternating molecular rows formed after annealing to \SI{350}{\celsius} ($U = \SI{-1.5}{V}$). (b) Overlay of molecular model on the structure ($U = +\SI{1.5}{V}$, $\SI{6.5}{nm}\times \SI{6.5}{nm}$). (c) Image of the occupied molecular orbitals at the left side of the ordered structure shown in (a). The green circle marks a lattice defect. (d) Defect along molecular rows (marked blue) from the incomplete C--C coupling between all molecules. (e) Model of the arrangement with covalent bonds between the debrominated molecules (red circle). Incomplete polymerization is marked by the green circle. Dissociated hydrogen atoms from the phenyl groups could diffuse and bond to the still unsaturated carbon atoms (arrows). Upwards directed C atoms in the pyrrole units of the porphyrin saddle-shape are colored in bright and dark yellow to indicate slightly different angles as explain in the text.}
\label{fig:struc_poly}
\end{figure*}

In the following, the observations in STM of the changes in the molecular arrangement after annealing will be discussed.
After deposition of a coverage of around one monolayer of CuTPPBr$_8$ on Au(111), many disordered areas but also small, periodically structured, close packed arrangements were found. Annealing of the sample drastically enlarged such ordered areas due to the increased mobility\cite{Buchner2011} of the molecules, which enables vast diffusion, rotation and, therewith, rearrangement and structure formation.
After heating to \SI{200}{\celsius}, most of the surface area was covered with the self-assembled molecular pattern, however, with still disordered regions between ordered domains. Fig.~\hyperref[fig:struc]{\ref*{fig:struc}(a)} shows two domains of the molecular structure which are rotated by \SI{120}{\degree} relative to each other due to the symmetry of the underlying Au(111) substrate. The "herring-bone" pattern of the $22\times\sqrt{3}$ reconstruction of Au(111) was visible through the molecular layer. This enables the direct measurement of the rotation of the molecular lattice relative to a substrate lattice vector, which is found to be $\varepsilon(\vec{A}) = \SI{45 \pm 5}{\degree}$.
At a bias of $U = \SI{-1}{V}$ (occupied states), the molecules appeared in an elongated shape [Fig.~\hyperref[fig:struc]{\ref*{fig:struc}(b)}], which reveals that the orientation of the molecules is rotated alternately by \SI{90}{\degree} in a checker-board-like manner. Such an appearance is characteristic for various metallo-tetraphenylporphyrin (TPP) molecules due to electron density at the site of the central metal ion and the well-known saddle-like deformation of the porphyrin macrocycle when adsorbed on surfaces.\cite{Scudiero2000,Buchner2010,Auwaerter2010,THPPstruc}
The distances between the molecules were measured to be $A = B = \SI{1.3 \pm 0.1}{nm}$, the angle of the unit cell $\alpha = \SI{88 \pm 2}{\degree}$ and the angle between a lattice vector and the orientation of the molecular saddle shape $\delta = \SI{10 \pm 5}{\degree}$. The model of the supramolecular arrangement derived from these values is shown in Fig.~\hyperref[fig:struc]{\ref*{fig:struc}(c)}. The arrangement is dominated by the attractive interaction of the edge-to-face configuration of adjacent phenyl groups\cite{Sinnokrot2004} which is typical for the self-assembly of various TPP on noble metal surfaces.\cite{Scudiero2000,Yoshimoto2003,Buchner2010,Auwaerter2010}
However, in these reported structures, the azimuthal orientation is usually the same for all TPP molecules which is not the case for the debrominated CuTPP molecules in the arrangement described here.

The XPS results indicate that after heating to \SI{200}{\celsius}, all CuTPPBr$_8$ molecules should have been debrominated and thus be surface-stabilized radicals. Despite that, the molecules didn't polymerize but formed a non-covalent structure (Fig.~\ref{fig:struc}) instead. This behavior can be understood by taking into account that, due to the molecular geometry, a C--C coupling between unsaturated pyrrolic carbon of neighboring molecules is strongly hindered by the phenyl groups. In this sense, after cleaving of the C--Br bonds, the reactive sites are protected against immediate polymerization because of steric reasons. The monomers can, thus, still freely diffuse and rearrange, which is highly desired for the formation of long range-ordered, close-packed networks with a small amount of defects.

After the sample was annealed to around \SI{350}{\celsius} for \SI{1}{h}, STM imaging revealed that the molecular lattice completely transformed into a new structure, which is shown in Fig~\ref{fig:struc_poly}. This new molecular network consists of apparently alternating rows of molecules with slightly different appearance, thus, adjacent molecules are not equivalent anymore. The distance between neighboring molecules is similar to the previously discussed structure in Fig.~\ref{fig:struc}. However, all molecules are rotated relative to each other when compared to the arrangement observed for annealing to max. \SI{200}{\celsius} -- the elongated shapes at $U = \SI{-1.5}{V}$ in one type of row are now oriented along the direction of the row [Fig.~\hyperref[fig:struc_poly]{\ref*{fig:struc_poly}(a)}]. The model of the newly formed molecular structure according to the measured distances and angles is shown in Fig.~\hyperref[fig:struc_poly]{\ref*{fig:struc_poly}(e)} and overlayed on the STM image in Fig.~\hyperref[fig:struc_poly]{\ref*{fig:struc_poly}(b)}.
A \SI{90}{\degree} rotation of adjacent molecules avoids cross-like overlapping of phenyl groups and results in favorable parallel-displaced $\pi$-$\pi$ stacking.\cite{Sinnokrot2004,Grimme2008}
The phenyl groups of adjacent molecules point to and are very close to the C atoms of the macrocycle where the bond to Br was cleaved. Thus, intermolecular covalent C--C bonds between an unsaturated pyrrolic C atom and a phenyl group were formed and the molecules were linked together. To make this possible, one C--H bond per phenyl group has to be cleaved.\cite{Sun2014,Palma2014} It is very likely that this dissociated hydrogen atom has been transfered inside the molecule along the phenyl group to one of the still unsaturated C sites as illustrated in Fig.~\hyperref[fig:struc_poly]{\ref*{fig:struc_poly}(e)}, which will later be discussed further.
The different appearance at the same bias voltage of every second molecule (``rows'') can be explained by the alternating orientation of the deformed macrocycle and respective bonding of phenyl to pyrrole units bended toward or away from the Au(111) surface. For every second molecule the pyrrole rings close to the substrate have to be bended up or the protruding pyrrole units slightly downwards toward the average molecular plane to minimize the stress on the new C--C bond with a phenyl ring of another molecule. Thus, the resulting saddle-shape deformations are non-equivalent for adjacent molecules, which was observed in STM [Fig.~\hyperref[fig:struc_poly]{\ref*{fig:struc_poly}(a)}]. A fusion of the pyrrolic C with the phenyl groups of the same molecule \cite{Shen2006, Roeckert2014} can be excluded in this case because this would result in the complete disappearance of a saddle-shape.

Each polymeric network covers a relatively large area. Nonetheless, coupling defects, which are typical for covalent networks, are still a big problem and will, thus, be discussed in the following. Four types of structural defects can be found. First, there are missing molecules creating holes in the network, which also occurs in the non-covalently bonded arrangement. On the lower left end of the ordered molecular structure in Fig.~\hyperref[fig:struc_poly]{\ref*{fig:struc_poly}(a)}, two rows with molecules showing the elongated shape are next to each other and also not displaced along the direction of the row as usual. The STM image of the unoccupied orbitals [Fig.~\hyperref[fig:struc_poly]{\ref*{fig:struc_poly}(c)}] shows a high density of states between each pair of neighboring molecules, which is an indication for a C--C coupling between the unsaturated carbon atoms of two pyrrole groups.\cite{Grill2007} In the same way, two molecules inside the same row [green circle in Fig.~\hyperref[fig:struc_poly]{\ref*{fig:struc_poly}(c)}] can be found closer together and link directly both porphyrin macrocycles. However, as discussed previously, for this a strong, energetically unfavorable deformation by bending the phenyl groups away would be necessary. The third type of defect, which occurs the most often, is shown in Fig.~\hyperref[fig:struc_poly]{\ref*{fig:struc_poly}(d)} and can also be seen magnified in Fig.~\hyperref[fig:struc_poly]{\ref*{fig:struc_poly}(b)}: The molecules in every second row can have different shapes which are related to bonding with all phenyl groups to neighboring molecules (``close-packed'' appearance) or C--C coupling with only two phenyl groups (slightly displaced and more elongated molecular shape). Defects should be problematic for applications in devices using lateral electronic transport, but at least the last type should be possible to cure by further annealing to promote complete C--C coupling between all neighboring molecules.

\subsection{DFT calculations}

\begin{figure*}[!tb]
\centering
\includegraphics[width=0.85\textwidth]{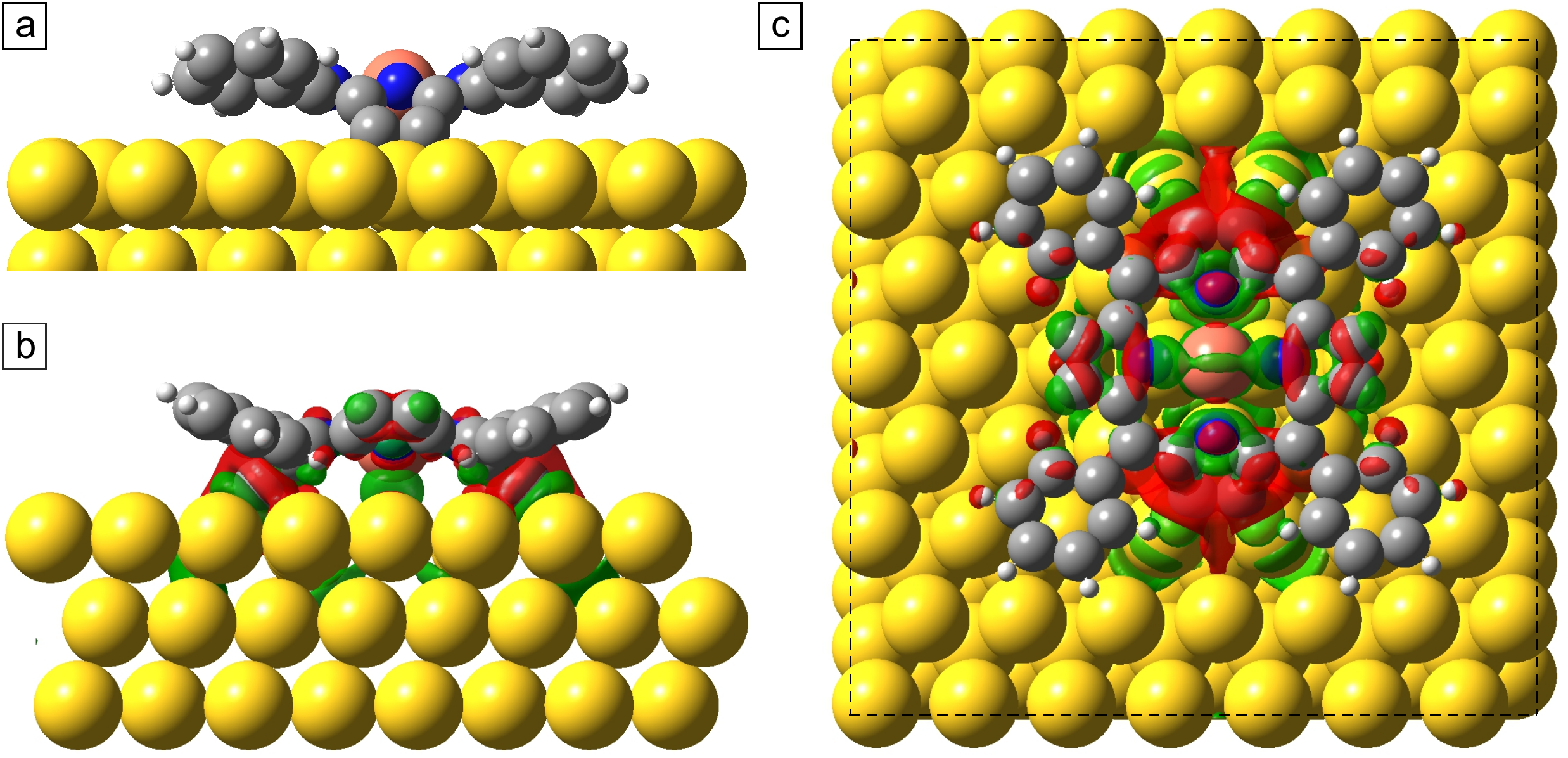}
\caption{(a) Optimized geometry of debrominated CuTPPBr$_8$ on a bridge adsorption site of Au(111). Isosurface of the difference of the total electron density between the combined CuTPP/Au system and the individual parts of molecule and Au slab in a side view (b) and top view (c). An increase of charge density is colored green, a depletion in red.}
\label{fig:DFT_CuTPP_Au}
\end{figure*}

In this section, we present DFT calculations [RPBE + $U$(\SI{5}{eV}) + vdW(TS)] of the debrominated CuTPPBr$_8$ molecules and of the polymeric structure.
In Fig.~\ref{fig:DFT_CuTPP_Au}, the optimized geometry is shown of a debrominated CuTPPBr$_8$ molecule (CuTPP) adsorbed on the favorable position with the Cu atom of the molecule \SI{2.8}{\angstrom} above a bridge position of Au(111). After the relaxation, the molecule has average angles of the phenyl planes of $\phi\_{ph} \approx \SI{18}{\degree}$, and of the pyrrole planes of $\phi\_{pyr}^{up} \approx \SI{19}{\degree}$ and $\phi\_{pyr}^{down} \approx \SI{37}{\degree}$ relative to the surface plane. Two pyrrole rings are, thus, strongly bent toward the surface to create C--Au bonds which stabilize the radical molecule, while the other pair of pyrrole rings remain with unsaturated carbon atoms.
The strong saddle-shape deformation is responsible for the elongated appearance in Fig.~\ref{fig:struc}.
The deformation energy to lift up the pyrrole rings which are bonded to the Au surface was calculated to be $\approx \SI{2.6}{eV}$. The necessity of an additional heating step for the polymerization could be due to the breaking of C--Au bonds and reversal of the deformation of the pyrrole rings and, thus, reactivation of these unsaturated carbon atoms. Furthermore, rotational and diffusion energy barriers have to be overcome for the reorganization of the molecular arrangement. Also, a high annealing temperature could be required due to bromine atoms which could be located between the molecules after debromination (although not visible in STM) which might hinder the polymerization until bromine is desorbed from the surface.

We also investigated the redistribution of the electron density upon the chemisorption of the molecule on the surface. The difference of the total electron density between the combined CuTPP/Au system and the individual parts of molecule and Au slab is shown in Fig.~\hyperref[fig:DFT_CuTPP_Au]{\ref*{fig:DFT_CuTPP_Au}(b),(c)}. The green colored areas correspond to an increased and the red areas to a decreased charge density. Especially at the C atoms which form the C--Au bonds, the total electron density is slightly reduced, whereas between the Cu atom and the surface, electron density increased and a small charge transfer of 0.1 electron to the Cu ion is found with a Bader charge analysis. This indicates also a direct interaction of the Cu ion in the chemisorbed molecule with the surface atoms due to the reduced distance after deformation of the molecule.

\begin{figure}[!tb]
\centering
\includegraphics[width=0.48\textwidth]{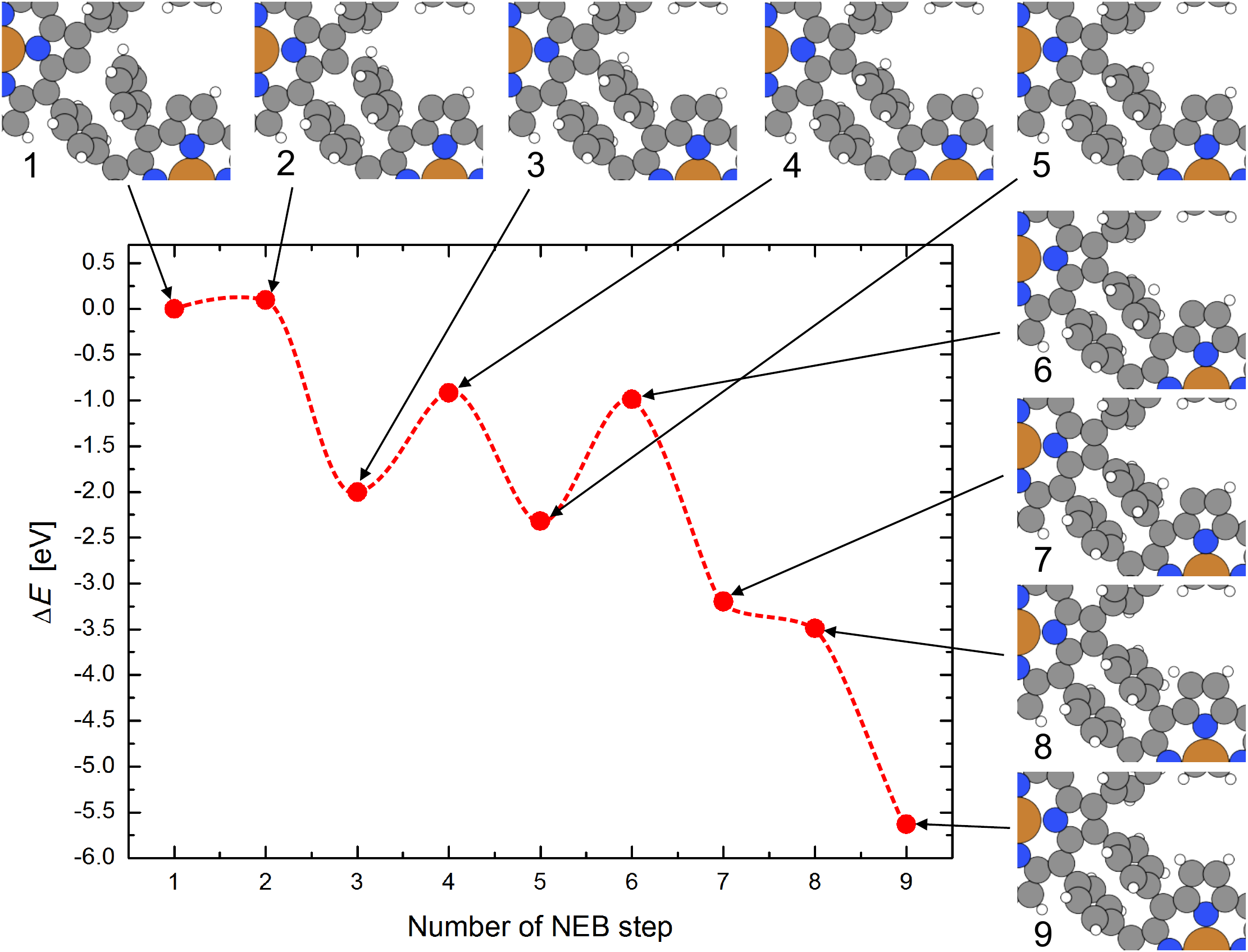}
\caption{Nudged elastic band path for the polymerization reaction between unsaturated CuTPP molecules in gas phase and the dislocation of the H atom from the newly formed C--C bond to the unsaturated pyrrolic C atom. The change of the atomic positions at each step is shown and linked by arrows to the corresponding change of the total energy.}
\label{fig:DFT_neb}
\end{figure}

Next, we calculated the structure of the CuTPP polymer in gas phase with the two coupled molecules in the periodic unit cell. The substrate could not be included in these calculations because the polymeric network is not commensurable with the Au lattice and the very large simulation cell which would be necessary for the coincidence of both unit cells was not computationally feasible for us. The dimensions of the unit cell were optimized by comparing the total energy of the polymer for different cell parameters after relaxation. The optimized size of the square unit cell was $x=\SI{2.145}{nm}, y=\SI{1.485}{nm}$, which is close to the experimentally measured values of \SI{2.3\pm 0.2}{nm} and \SI{1.56\pm 0.20}{nm}.
The calculated angle of the covalently connected pyrrole group of $\phi\_{pyr}^{poly} \approx \SI{7}{\degree}$ is reduced compared to the angle of the not coupled pyrrole with $\phi\_{pyr} \approx \SI{15}{\degree}$ in gas phase, which results in the non-equivalent appearance for adjacent molecules in STM as previously discussed.

To investigate the intramolecular displacement of a hydrogen atom from the phenyl group to an unsaturated carbon atom in the pyrrole unit after the coupling, a ``nudged elastic band'' (NEB) calculation with 8 steps was performed, whereby the molecules in the periodic cell are already covalently coupled except at one phenyl and pyrrole location. The calculated reaction path is shown in Fig.~\ref{fig:DFT_neb}. First in step 1 to 3, the phenyl group bends to get closer to the adjacent debrominated pyrrole group, whereby the H atom has to move out of the plane of the phenyl ring to enable the formation of the new covalent C--C bond between the unsaturated C atom and the phenyl ring. For this coupling reaction an energy barrier of below \SI{0.1}{eV} was found (in gas phase). The polymerization could, thus, occur directly after debromination if the molecules are close enough and correctly oriented. Starting from NEB step 3, the H atom jumps on a path along the C atoms of the phenyl group to the still unsaturated pyrrolic C atom. The maxima of the energy barriers for the jumps correspond to the position of the H atom exactly between two adjacent C atoms. The relocation of hydrogen to the still unsaturated C atoms leads to a large gain in energy, thus, it is very likely that at least all pyrrolic C atoms of the CuTPP-polymer which are bent up and not in direct contact with the Au surface are saturated with H atoms.

\section{Conclusions}

The heat-induced coupling-limited polymerization of CuTPPBr$_8$ on Au(111) by dissociation of bromine was studied by XPS and STM, and the results are corroborated with DFT calculations.
For annealing until \SI{200}{\celsius}, CuTPPBr$_8$ molecules debrominate and arrange in a square structure where the azimuthal orientation of the molecules is rotated by \SI{90}{\degree} in a checker board-like manner. Upon heating of the molecular monolayer to \SI{300}{\celsius} or higher, C--C coupling between C atoms of the phenyl rings and unsaturated C in the pyrrole units of adjacent molecules is initiated, including the cleavage of a C--H bond at each phenyl group. Consequently, the molecules form a relatively large, covalently bonded network. Interestingly, shake-up peaks in XPS indicate that thereby the HOMO--LUMO gap opened by \SI{0.4}{eV} instead of a decrease, which shows that the conjugation is not extended. The porphyrin macrocycles can, thus, still be considered as individual intact units although they are bonded covalently in a large robust 2D network. Extended two-dimensional polymeric sheets with high chemical and mechanical stability, especially from building blocks which include also magnetic metal ions could be promising for spintronic devices. Also, the lateral transport properties could be interesting for organic interconnections in novel molecular electronic devices. For this, however, it has to be possible to remove the polymer from the metallic surface or to decouple it, e.g. by a thin insulating buffer layer, which will be a big challenge.
A next step is also to investigate the influence of the diffusion for the polymerization of CuTPPBr$_8$ further by choosing more reactive surfaces with a higher diffusion barrier, which we will describe in a following publication.

\section*{Acknowledgements}

This work has been financially supported by the Deutsche Forschungsgemeinschaft (DFG) through the Research Unit FOR 1154 and the Fonds der Chemischen Industrie. Photoelectron spectroscopy was performed at the Material Science beamline at the synchrotron Elettra (Trieste, Italy). We thank Martin Vondr$\acute{\mathrm{a}}$\v{c}ek for technical assistance. M.K. thanks the Fonds der Chemischen Industrie for a Chemiefonds fellowship. Computational resources were provided by the "Chemnitzer Hochleistungs-Linux-Cluster" (CHiC) at the Technische Universität Chemnitz.

\footnotesize{
\bibliography{CuTPPBr8_Au}
\bibliographystyle{rsc} 
}

\end{document}